\documentclass[sigconf]{acmart}

\AtBeginDocument{%
  }

\copyrightyear{2026}
\acmYear{2026}
\setcopyright{cc}
\setcctype{by}
\acmConference[WebSci '26]{18th ACM Web Science Conference}{May 26--29, 2026}{Braunschweig, Germany}
\acmBooktitle{18th ACM Web Science Conference (WebSci '26), May 26--29, 2026, Braunschweig, Germany}
\acmDOI{10.1145/3795766.3799753}
\acmISBN{979-8-4007-2504-3/2026/05}

\usepackage{multirow} 
\usepackage{array} 
\usepackage{subcaption}
\usepackage{tabularx}

\usepackage{booktabs} 
\usepackage{newfloat}
\usepackage{listings}

\usepackage{pifont}

\usepackage{makecell} 
\usepackage{lscape}

\usepackage{enumitem}

\begin{document}


\title{Why They Link: An Intent Taxonomy for Including Hyperlinks in Social Posts}

\author{Fangping Lan}
\email{fangping.lan@temple.edu}
\affiliation{%
  \institution{Temple University}
  \city{Philadelphia}
  \state{PA}
  \country{USA}
}
\author{Abdullah Aljebreen}
\email{az.aljebreen@su.edu.sa}
\affiliation{%
  \institution{Shaqra University}
  \city{Shaqra}
  \country{Saudi Arabia}}

\author{Eduard Dragut}
\email{edragut@temple.edu}
\affiliation{%
  \institution{Temple University}
  \city{Philadelphia}
  \state{PA}
  \country{USA}
}


\begin{abstract}
URLs serve as bridges between social media platforms and the broader web, linking user-generated content to external information resources. On Twitter (X), approximately one in five tweets contains at least one URL, underscoring their central role in information dissemination. While prior studies have examined the motivations of authors who share URLs, such author-centered intentions are difficult to observe in practice. To enable broader downstream use, this work investigates reader-centered interpretations—how users perceive the intentions behind hyperlinks included in posts. We develop an intent taxonomy for including hyperlinks in social posts through a hybrid approach that begins with a bottom-up, data-driven process using large-scale crowdsourced annotations, and is then refined using large language model (LLM) assistance to generate descriptive category names and precise definitions. The final taxonomy\footnote{Our complete intent taxonomy is publicly released at \url{https://github.com/lanfangping/intent-taxonomy-for-hyperlinked-posts}} comprises 6 top-level categories and 26 fine-grained intention classes, capturing diverse communicative purposes. Applying this taxonomy, we annotate and analyze 1,000 user posts, revealing that advertising, arguing, and sharing are the most prevalent intentions. We further compare our taxonomy with existing taxonomies and demonstrate its utility in a microblog retrieval task by incorporating intent as an additional feature. Overall, our taxonomy provides a foundation for intent-aware information retrieval and NLP applications, enabling more accurate retrieval, recommendation, and interpretation of social media content.

\end{abstract}




\maketitle

\section{Introduction}

Communication scientists have developed several research agendas seeking to address the long-standing question ``What do people do with media?'' \cite{Katz1959Mass}. 
In traditional mass communication (e.g., television, radio, newspapers), this inquiry gave rise to the uses and gratifications research tradition \cite{Ruggiero2000Uses}, an audience-centered framework explaining how individuals engage with media to satisfy specific needs such as information seeking, entertainment, or social connection \cite{Holbert2014Uses, Haridakis2019Uses, https://doi.org/10.1002/widm.1342}. From an Information Retrieval (IR) perspective, these needs can be interpreted as information-seeking behaviors, where media use serves as a way of discovering, consuming, or disseminating information objects.
With the rise of social media platforms, users are no longer only consumers of content but also content creators \cite{Chen2011Tweet, Phua2017Uses, Smock2011Facebook}. This shift implies that users simultaneously act as query issuers and document providers.

Social media posts contain not only textual content but also structural and semantic elements such as hashtags, emojis, images, and hyperlinks (or URLs) \cite{Yang2012We, Vempala2019Categorizing, Hu2017Spice}. Existing research has examined social media content at different levels, including general tweet intent \cite{Java2007Why}, topic-focused messages \cite{Mohammad2015Sentiment}, and component-specific studies such as hashtags or emojis \cite{Hu2017Spice}. 
It has been shown that hashtags, emojis, and images can function as rich indexing and query-expansion features in social media search (aka, microblog retrieval) \cite{wang-2023-kdd, moumita-2018-web}  and recommendation tasks \cite{hashtagsuggestion, emojisearch, AReviewoftheStudiesonSocialMediaImages}. However, URLs, which directly link social media posts to external web documents, have received comparatively less attention.
Our prior work established that tweets containing URLs
form a distinct and computationally important class of posts requiring dedicated modeling \cite{Aljebreen2021Segmentation}. 
Yet, URLs carry particularly strong retrieval implications: they act as pointers to external information resources and encode users’ intentions for exposing or consuming content. Understanding why users include URLs is important for a range of IR tasks, from intent-aware microblog retrieval ranking \cite{Cambazoglu2021An, li2023intentawarerankingensemblepersonalized} to credibility assessment and crisis informatics applications \cite{purohit2014identifying, Tanaka2013The, Horawalavithana2021Malicious}. 

\begin{figure}[!t]
  \centering
    \includegraphics[width=0.99\columnwidth]{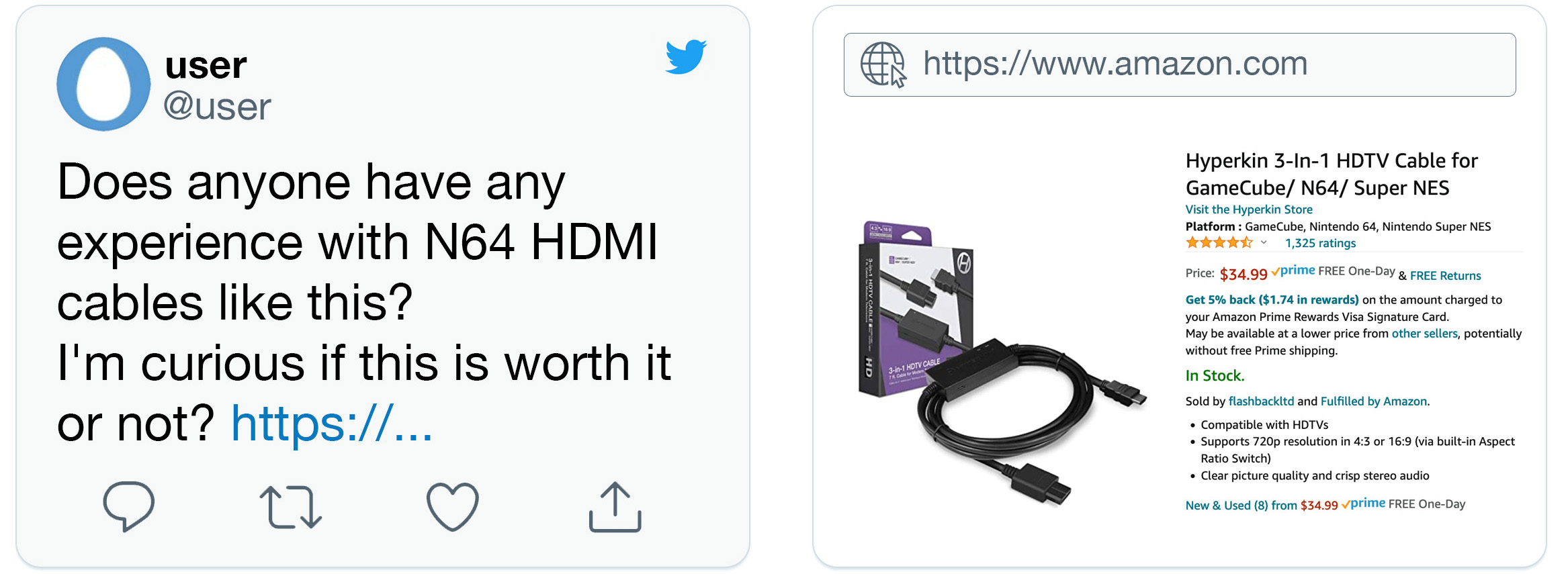}
    \vspace{-1em}
  \caption{An example of a tweet with the URL of a product web page, with the intent of inquiring about it.}
  \label{fig:t1}
\end{figure}
\begin{figure}[!t]
  \centering
    \includegraphics[width=0.99\columnwidth]{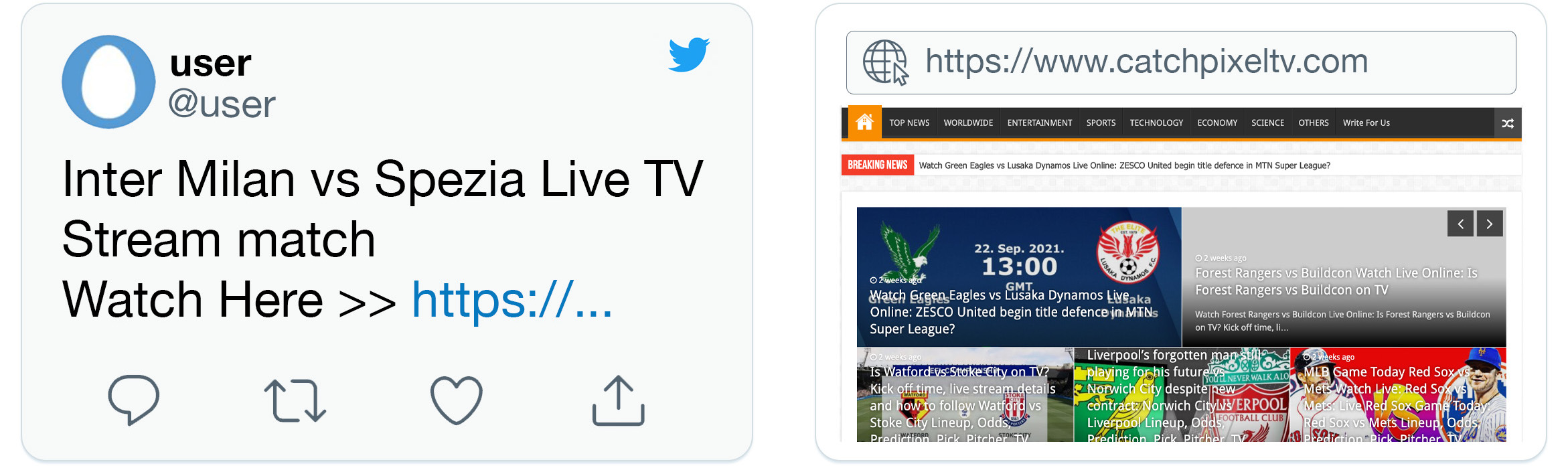}
    \vspace{-1em}
  \caption{An example of a tweet that promises a free streaming for a sport match through the URL.}
  \label{fig:t2}
\end{figure}

Our goal is therefore to construct a comprehensive, empirically grounded taxonomy of user intentions behind URL sharing on microblog platforms, such as Twitter (X). Prior studies have primarily explored this question from the authors’ perspective, using surveys or interviews to examine their motives for including URLs in posts \cite{Holton2014Seeking, BAEK20112243}. While such approaches capture authentic creator intentions, they are difficult to generalize and impractical for large-scale application, as authors’ true motives are rarely observable in real-world settings. \textit{We instead focus on intentions from the perspective of readers—that is, how other users interpret the purpose of including a URL.}
Inference of URL-sharing intention is not straightforward. The meaning of a URL often cannot be deduced without considering both the post’s text and the linked content.
For example, a post that links to an Amazon product web page may appear promotional if we ignore the post's text (Figure \ref{fig:t1}). However, the text of the post reveals that the intention is not to advertise the product but rather to solicit the readers' feedback about it. 
In another case, a post may claim to provide free streaming for a sporting event, yet the URL leads only to a news site, exposing the latent intention as clickbait (Figure \ref{fig:t2}).
The latter is an example 
promoting affiliated websites \cite{Cable2018Can, Chakraborty2017Tabloids}. 
These examples highlight the need for an intent taxonomy for hyperlinked posts. A well-defined taxonomy can provide a foundation for intent-aware indexing, ranking, recommendation, and credibility assessment \cite{ORCAS-I-2022, Cambazoglu2021An, Pasi2020InformationCI}. By distinguishing between intentions such as information sharing, promotion, discussion, or deception, IR systems can better model relevance, improve user experience, and support downstream applications such as credibility detection and crisis response \cite{AIDR2014, Wang-Misinformation-2024}. 

The main objective of our work is to examine the underlying intentions behind including hyperlinks in social posts from a reader-centered perspective. We address two tasks in this work: (1) Develop a comprehensive taxonomy that categorizes the perceived intentions of posting URLs in posts; (2) Study and analyze the intention distribution and impacts of annotation reliability through two user studies involving a random sample of real-world hyperlinked posts.

The structure and main contributions of our work are as follows:
\begin{itemize}
    \item 
    Section \ref{sec:motivation_sec} discusses the importance of understanding readers' interpretations of intentions behind including URLs in posts and describes how a well-structured taxonomy enables systematic investigation of these interpretations.
    \item Section \ref{sec:taxonomy_process} presents our process of developing an intent taxonomy for categorizing the intentions behind URLs in social posts. It leverages a hybrid approach that combines human annotation and LLM-based assistance. The resulting taxonomy comprises 6 broad intention categories and 26 specific intention classes. For simplicity, we use the term intention class to refer to both levels.
    \item Section~\ref{sec:labeling} presents two user studies that examine how users interpret the intentions behind URL-sharing posts. The first study analyzes the distribution of intentions for including URLs across a large collection of tweets. The second study examines how annotation quality changes when expert annotators are involved and additional contextual information is provided, aiming to improve the labeling quality of low-agreement posts.
    \item Section \ref{sec:evaluation} compares our intent taxonomy with those proposed in prior work and presents a potential application in microblog retrieval, where intent information is incorporated as an additional feature for reranking and filtering retrieval results.
\end{itemize}

\section{Motivation}
\label{sec:motivation_sec}
Among the many components of social posts, URLs stand out 
because they connect social media posts to external information resources. From an IR perspective, URLs function not merely as enrichment to the message but as surrogates for external documents, linking user-generated content to the broader web. Their prevalence is significant: approximately one out of five tweets contains at least one URL \cite{Aljebreen2021Segmentation}, and over half of retweets involve URL-linked tweets \cite{Suh2010Want}. This frequency highlights URLs as central carriers of information flow between Twitter and the web. Understanding the intentions behind including URLs in posts is thus a crucial step toward intent-aware retrieval, credibility assessment, and recommendation. In the remainder of this section, we examine the scope of existing taxonomies and highlight the utility of a URL-intention framework for IR applications.

\subsection{Scope and Gaps in Existing Taxonomies} \label{sec:gap}
Prior work has proposed several taxonomies for categorizing tweet-level intentions \cite{Java2007Why, Alhadi2011Exploring, GmezAdorno2014Content}. However, these frameworks were designed to classify the communicative goals of posts as standalone messages and do not account for the distinct role of URLs as anchors to external documents. Our pilot study confirmed this gap: we hired two graduate students to annotate 200 URL-linked tweets using categories compiled from \cite{Alhadi2011Exploring, GmezAdorno2014Content}, 32.5\% were labeled as \textit{None/Other} because no suitable category existed. In addition, some categories, such as \textit{Chat} \cite{GmezAdorno2014Content}, were never applied, indicating their irrelevance for URL-driven intentions.  

These limitations are significant for downstream applications. Current taxonomies lack clear definitions and representative examples, making them difficult to apply consistently and unsuitable for large-scale annotation. More importantly, they overlook key distinctions among intentions such as genuine information sharing, promotion, or deceptive link posting—differences that are vital for intent-aware retrieval and ranking systems \cite{Cambazoglu2021An, li2023intentawarerankingensemblepersonalized}. 

\subsection{The Utility of a URL-sharing Intention Framework}
An intent taxonomy for including hyperlinks in posts offers structured signals that can improve retrieval and recommendation systems across multiple domains. We highlight three key application areas below.

\textbf{Crisis Response}\label{sec:cr}
In the domain of crisis response and emergency awareness, extensive research has been conducted that leverages posts and other forms of social media content \cite{pipek2014crisis}. Several studies pay attention to the presence of URLs in crisis-related tweets, with some reporting that over 50\% of the tweets in their datasets contain URLs \cite{purohit2014identifying,Hughes2009Twitter, Meesters2016Help, Hunt2020Misinformation}. 
For IR systems that support crisis informatics, distinguishing URL-sharing intentions is critical. A taxonomy enables retrieval pipelines to prioritize actionable links (e.g., shelter information, emergency hotlines) while downranking malicious or irrelevant ones. It also aids in credibility assessment by signaling whether URLs are shared to inform, to seek help, or to mislead \cite{Horawalavithana2021Malicious}. Such intent-aware retrieval can improve the timeliness and accuracy of information accessible to crisis responders. 

\textbf{Misinformation}
Our prior work demonstrates that aggregating social media posts by shared URLs reveals structural regularities and dissemination risks that remain invisible at the post level, enabling highly accurate detection of deceptive link-sharing patterns \cite{AljebreenMD24}. Building on this insight, the proposed URL intent taxonomy provides a principled framework for systematically characterizing the motivations behind URL-sharing behaviors. By categorizing the intentions underlying link dissemination, it supports more precise detection of malicious or deceptive campaigns, facilitates retrospective analyses of misinformation propagation, and enables proactive interventions grounded in credible information dissemination. Thus, the taxonomy contributes to promoting trustworthy content and fostering a healthier digital information ecosystem.


\textbf{NLP and IR Tasks}\label{sec:NLPIR}
In many NLP tasks, such as sentiment analysis, URLs are often treated as peripheral and removed from tweets without exploiting their potential signals \cite{Chong2014Natural, Shivakumar2021A}. We argue that a taxonomy of URL-sharing intentions enables these signals to be reintroduced as interpretable features for both NLP and IR applications. The insights derived from URL-sharing intentions offer signals of the communicative purpose, quality, and utility of a post. For instance, distinguishing whether a URL is included to promote, to inform, or merely to reference can enhance sentiment analysis \cite{WangDragut2024RLF,Dragut10}, stance detection \cite{Hosseinia19}, and user modeling \cite{languageIntentModels}. From an IR perspective, intent-aware representations support richer query understanding, improve document (microblog) ranking, and lead to more precise retrieval \cite{Cambazoglu2021An, li2023intentawarerankingensemblepersonalized,Zhang2020BirdsFeather}. 
As our demonstration reports (Section \ref{sec:example_retrieval}), 
recognizing intention allows retrieval systems to better estimate which posts are valuable for a given query, thereby increasing confidence in their relevance.

\section{Developing the Intent Taxonomy}
\label{sec:taxonomy_process}

The large fraction of social media posts containing URLs \cite{Aljebreen2021Segmentation, Suh2010Want} highlights the diversity of content being circulated and suggests a wide range of underlying intentions behind URL sharing. We describe our approach for constructing a tailored taxonomy that captures a comprehensive set of perceived intentions behind URL inclusion in posts in this section. Further details are available in the supplementary material.




\subsection{Data and Participants}
The process of constructing the taxonomy involves crowdsourced annotations collected from the Amazon Mechanical Turk (AMT) platform with Twitter data. In this section, we give the details of the dataset and the participants in the annotation process.

\subsubsection{Dataset}
Our \textit{Tweets with URLs} dataset consists of a random collection of tweets, which was collected using our previously designing API-driven and focused crawling pipelines for large-scale social media and web research~\cite{ChenHMD24,ChenMD25,AljebreenMD24,Aljebreen2021Segmentation}. We deliberately avoid using existing tweet datasets, as many of them suffer from one of two limitations: they either remove URLs during normalization or retain URLs that are no longer accessible, often redirecting to error pages or generic homepages. To address these issues, we collect recent tweets and simultaneously crawl the linked web documents, increasing the likelihood of preserving valid tweet–URL pairs. Access to the linked content is essential for our study, as annotators are instructed to consult the URL content when the tweet text alone is insufficient to infer intent. The \textit{Tweets with URLs} dataset contains over 500,000 tweets. From this collection, we randomly sample 2,500 tweets for taxonomy development.


\subsubsection{Participants}
Given documented concerns about the decline in data quality on Amazon Mechanical Turk (AMT) in recent years— particularly the rise of low-quality and low-fluency responses~\cite{Marshall2022WhoBrokeAMT}— we adopted a stricter worker screening process to improve the reliability of crowdsourced annotations. We employed a qualification task to screen AMT workers. The task consisted of annotating five carefully selected tweets and was restricted to AMT Master workers. Workers were evaluated on their understanding of the task and their ability to identify the intended purpose of URL sharing. Out of the 200 workers who completed the qualification task, 143 met our acceptance criterion by correctly annotating at least four tweets and were marked as \textit{eligible}. Only \textit{eligible} workers were allowed to participate in the subsequent taxonomy development.


\subsection{Procedures and Results}
\begin{figure}[!t]
  \centering
\includegraphics[width=0.9\columnwidth]{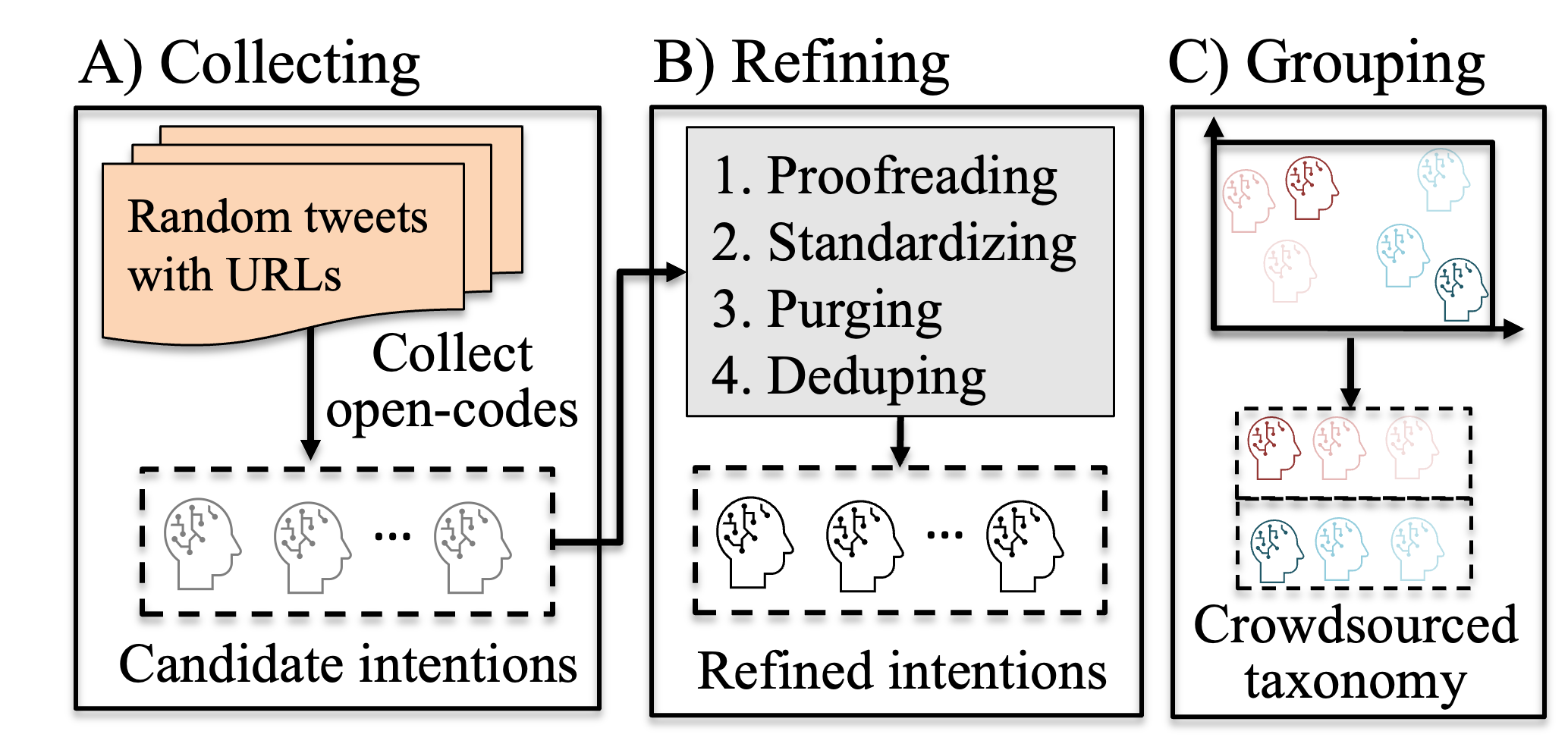}
\vspace{-1em}
  \caption{Our 3-phase framework of developing our intent taxonomy for sharing URLs in Twitter}
  \label{fig:framework-tax}
 \end{figure}

Our intent taxonomy is built upon the insights of previous research. We followed the approach of \cite{Talevich2017Toward} in developing our framework for intent taxonomy. Our framework consists of three steps: (A) collecting a list of candidate intentions, (B) refining the intentions, and (C) grouping them into categories based on similarity. The process is illustrated in Figure \ref{fig:framework-tax}.

\textbf{(A) Collecting.}
We began our study with a crowdsourcing task on AMT, where participants labeled tweets with their perceived intentions behind URL inclusion. Crowdsourcing provides a scalable and cost-effective solution for gathering a large volume of data from diverse individuals \cite{Cocos2017Crowd, Bloodgood2010Using}. We opted for an open-coding approach, allowing AMT workers to provide codes or intention classes for tweets using an open-text form. Open coding is a popular method in qualitative research and grounded theory which has been used in several works to construct taxonomies in other problems \cite{Scott2012Adapting}. This method allowed us to capture a wide range of potential intention classes from the crowd and ensured that workers had the flexibility to express their thoughts and observations without being restricted by predefined categories. 

In our study, we engaged the participation of 25 $eligible$ AMT workers to annotate a set of 100 tweets each.  
By distributing the task across multiple workers, we aimed to capture a diverse range of perspectives and interpretations of intentions. The diversity of annotations is valuable as it provides a broader understanding of the intentions in tweets with URLs, and provides depth to the desired taxonomy. By incorporating multiple annotations from different workers, we can account for individual biases, preferences, and viewpoints, thereby ensuring a more comprehensive and well-rounded taxonomy. After gathering the initial set of codes we proceeded to the next step which was data processing and cleaning.

\textbf{(B) Refining.} We refined 2,500 intention classes in the previous step into a final list of 754 intention codes. The process is shown as follows:

\textit{Proofreading:} We manually inspect the list for grammatical errors and typos and finally correct 90 codes. 

\textit{Standardizing:} We unify the phrasing of the intentions by phrasing them as answers to the question, \textit{“Why did you share this link?”}. For example, the code "\textit{sharing news article}" is paraphrased to "\textit{to share a news article}." By doing this we ensure the clarity of the intentions in the list.

\textit{Purging:} We performed a 
thorough filtering process to ensure the quality and reliability of the results. We specifically looked 
for unclear codes or ones that do not appear as actual intention codes but rather appear as descriptive statements or explanations.  For example, the code "\textit{talk and complain about rent increase}" is a statement about the tweet itself rather than an intention code about the URL, which in that case should have been ``to support an argument.'' 
Out of the 2,500 annotations in the list, we deemed 30 instances unsuitable and discarded those codes. We focus on the annotations that provided clear and concise intention codes.

\textit{Deduping: } We removed duplicate annotations from the list. We eliminated both exact duplicate codes and the semantically similar ones. For example, the code "\textit{to provide an answer to a question}" is similar to "\textit{to answer a question}." We also asked workers to tell us when two different intention codes had the same meaning. We describe that in the next step.

\textbf{(C) Grouping.} Given that we have a list of hundreds of intention codes, our next step is to convert it into an empirically derived hierarchical taxonomy with high-level intentions categories that represent the spectrum of intentions in our problem. We use \textit{affinity mapping} to refine the intentions list and produce the taxonomy we desire for our problem. Affinity mapping is a common research approach that is used to group gathered data based on their similarity and discover their underlying structure \cite{Otieno2023Affinity}. 

In this phase, we employed a group of 5 AMT workers to work on the grouping task. Out of the five workers, three participated in the initial annotation (Phase A). In this task, the workers were asked to sort the list of intentions into categories based on what went together. Each new category was  named by the worker. The workers were also asked to indicate if a pair of codes were identical in their meaning, but had different spelling. This allowed us to merge those that were semantically similar.

We used a simple majority to process the final categories and other decisions. Based on the majority votes, 311 were discarded given their similarity to other codes. In the grouping results, some responses formed more categories than others. The number of groups ranged from 17 to 31 in the responses. We compiled those groups into 28 intention groups with a total of 442 fine-grained intentions. After the initial grouping, we attempted to establish higher-level classes of intentions. To achieve this, we conducted another round of grouping with a new set of 5 AMT workers. As a result, we generated a shorter list of 6 coarse-grained intention classes (Table \ref{tab:final_taxonomy}).

\subsection{A Hybrid Construction Method}
\label{sec:hybrid-construction}
We developed our intent taxonomy for including URLs in posts using a \textit{three-phase, bottom-up, data-driven framework}. In this approach, categories are derived directly from real examples, ensuring that each class is grounded in authentic cases that enhance interpretability and user understanding. The bottom-up method excels at capturing fine-grained distinctions and providing concrete illustrations, which explains its widespread adoption in prior studies \cite{Java2007Why,Alhadi2011Exploring}. However, purely bottom-up taxonomies often suffer from vague category names and imprecise definitions. Conversely, \textit{top-down, experience-based} approaches construct taxonomies by aligning and merging categories from existing frameworks based on their names, definitions, and examples \cite{lan-etal-2025-unit}. This strategy yields clearly defined and well-labeled categories but tends to miss subtle distinctions and lacks empirical grounding, while also being time-consuming when performed manually.
Recently, LLMs have emerged as a promising complementary tool \cite{UsingLargeLanguageModels}: even without direct access to raw data, they can generate taxonomies that implicitly draw on prior knowledge, particularly excelling at proposing descriptive labels and coherent definitions. Building on the complementary strengths of these approaches, we adopt a hybrid method: first applying a bottom-up process to identify candidate categories and collect representative examples, and then using LLMs to refine the taxonomy by assigning descriptive names and precise definitions.


\begin{table*}[ht]
\centering
\caption{The intent taxonomy for including hyperlinks in posts. The taxonomy is constructed using a hybrid approach: bottom-up data-driven discovery of intention classes, refined with descriptive names, definitions, and illustrative examples. It contains 6 top-level categories and 26 intention classes. The category names in parentheses are from crowdsourcing. }
\small
\label{tab:revised-taxonomy}
\begin{tabular}{p{0.14\linewidth}|p{0.19\linewidth}|p{0.32\linewidth}|p{0.265\linewidth}}
\hline
\textbf{Category} & \textbf{Definition} & \textbf{Intention Classes} & \textbf{Illustrative Example} \\ \hline \hline

\multirow{8}{*}{\makecell[l]{\parbox{2cm}{Information \\ Sharing (Share)}}} & \multirow{8}{*}{\parbox{3.45cm}{Disseminating information, updates, or personal news.}} & 
To share political or public news & e.g., link to election coverage \\ \cline{3-4}
& & To share health-related news & e.g., COVID-19 guidelines \\ \cline{3-4}
& & To share economy, science, or technology news & e.g., new AI model release \\ \cline{3-4}
& & To share sport-related content & e.g., game highlights \\ \cline{3-4}
& & To share personal achievements & e.g., graduation photos, awards \\ \cline{3-4}
& & To share current activities & e.g., event attendance \\ \cline{3-4}
& & To share user-generated content & e.g., blog posts, YouTube videos \\ \cline{3-4}
& & To share personal information & e.g., personal websites, résumés 
\\ \hline  \hline

\multirow{2}{*}{\makecell[l]{\parbox{3cm}{Entertainment /\\ Humor (Entertain)}}} & \multirow{2}{*}{\parbox{3.45cm}{To provide amusement or entertainment content.}} & To share humorous content & e.g., memes, parody sites \\ \cline{3-4}
& & To share entertainment news & e.g., celebrity or movie-related articles  \\ \hline \hline

\multirow{4}{*}{\makecell[l]{Assistance /\\ Information Provision \\ (Offer)}} & \multirow{4}{*}{\parbox{3.45cm}{Offering help, advice, or useful information to others.}} & To provide feedback or answer questions & e.g., linking to articles in feedback\\ \cline{3-4}
& & To offer help or advice & e.g., tutorials, how-to guides \\ \cline{3-4}
& & To share personal experiences about a merchandise & e.g., product/service reviews \\ \cline{3-4}
& & To provide factual information & e.g., background links in debates  \\  \hline \hline
\multirow{4}{*}{\makecell[l]{\parbox{2.5cm}{Discussion /\\ Opinion Expression (Converse)}}} & \multirow{4}{*}{\parbox{3.45cm}{Using URLs to support or extend conversation, arguments, or emotions.}} & To support an argument & e.g., linking to evidence in debates \\  \cline{3-4}
& & To express or add context to an opinion & e.g., linking to explanations or answers \\ \cline{3-4}
& & To express emotions & e.g., frustration, excitement tied to a link \\ \cline{3-4}
& & To discuss or provoke discussion & e.g., controversial news links  \\ \hline \hline

\multirow{4}{*}{\makecell[l]{\parbox{2cm}{Promotion /\\ Advertisement (Promote)}}} & \multirow{4}{*}{\parbox{3.45cm}{Recommending or promoting products, events, or personal content.}} & To recommend content to others & e.g., ``Check this out! [link]'' \\ \cline{3-4}
& & To promote products or services & e.g., store links, affiliate marketing \\ \cline{3-4}
& & To promote events & e.g., concerts, conferences \\  \cline{3-4}
& & To promote personal or external content & e.g., blog posts, podcasts \\ \hline \hline

\multirow{4}{*}{\makecell[l]{\parbox{2cm}{Request /\\ Call for Action (Request)}}} & \multirow{4}{*}{\parbox{3.45cm}{Seeking feedback, engagement, or assistance from others.}} & To ask questions or seek feedback & e.g., ``What do you think of this? [link]'' \\  \cline{3-4}
& & To seek help or support & e.g., donation links, petitions \\ \cline{3-4}
& & To persuade others to visit & e.g., click-through bait without promotion \\ \cline{3-4}
& &  To seek connection or collaboration & e.g., networking, group invites \\ \hline

\end{tabular}
\label{tab:final_taxonomy}
\end{table*}


We present our intent taxonomy for including hyperlinks in social posts\footnote{The complete taxonomy with representative examples can be found at the shared link.} in Table~\ref{tab:final_taxonomy}.
The final taxonomy represents a synthesis of human annotation and LLM-based refinement. Specifically, we enhanced the taxonomy derived from crowdsourced annotations by prompting GPT-5 to propose more descriptive category names and clearer definitions, using the crowdsourced categories and fine-grained intent classes as input. The LLM-generated suggestions were then reviewed and consolidated through a consensus discussion among the research team to ensure interpretability, consistency, and empirical grounding.
We also sought to enrich each fine-grained intention class with both illustrative examples and example posts. Prior studies and our experience indicate that annotators prefer taxonomies that include examples alongside definitions when applying them to new annotation tasks \cite{HettiachchiKG22, pradhan2021searchambiguitythreestageworkflow}. Hence, we used LLMs to generate illustrative examples and revised them following the steps described above (see the  Illustrative Example column in Table \ref{tab:final_taxonomy}). 
We additionally explored using LLMs to create example posts for each intention class. However, this revealed a key limitation: while LLMs can generate plausible and often representative posts, the examples are synthetic rather than drawn from real-world data. Consequently, all examples in our final taxonomy were manually retrieved and curated from authentic social media posts for each of the 26 intention classes.

\section{Study: Crowd Labeling of URLs Intention}\label{sec:labeling}
The next step in our work is to employ the newly developed taxonomy in user annotation studies. In this section, we show and discuss the findings of these user studies,
which investigate how users interpret the intentions behind URL-sharing posts. First, we describe \textbf{Study 1}, which examines the overall distribution of intentions in a random set of tweets using our taxonomy. Next, we revisit a subset of the tweets from Study 1 where annotators did not reach high agreement. We conduct a follow-up study, \textbf{Study 2}, in an attempt to increase the agreement on these difficult instances. Across both studies, we employ the six top-level intention categories in our taxonomy (Table \ref{tab:final_taxonomy}).


Before proceeding, we define the terminology used in this section to describe the annotation outcomes for tweets:

\begin{itemize}[nosep, left=2pt]
    \item \textbf{High consensus}: at least four of the five annotators select the same intention class for a tweet.
    \item \textbf{Competing consensus}: three of five annotators agree on the same class; the other two agree on same other class.
    \item \textbf{Split decision}: aside from the three agreeing annotators, the two remaining annotators choose different intention classes.
    \item \textbf{No majority}: no intention class receives at least three votes.
\end{itemize}

If no intent category defined in our taxonomy is suitable for a tweet, the annotator may label it as \textbf{uncertain}.
Tweets labeled as either \textbf{uncertain} or with \textbf{no majority} are collectively referred to as \textbf{NC-UN}.

\subsection{Dataset} We randomly and uniformly sampled 1,000 tweets from our \textit{Tweets with URLs} dataset to reduce potential temporal bias and improve the robustness of our taxonomy. 
Table~\ref{tab:results-properties} summarizes key tweet characteristics, including type, length, and total reactions.
Tweet \textit{type} includes replies, quoted tweets, and other types of tweets.
For the \textit{reactions} attribute, we quantify tweet popularity using three engagement signals: \textit{likes}, \textit{replies}, and \textit{retweets}.
All metrics were collected more than 24 hours after publication and are assumed to reflect stable engagement levels \cite{pfeffer2023half,He2021CannotPredict}.


\begin{table}[!t]
    \caption{Distribution of the top three intention classes across tweet properties (type, length, and reaction count).}
    \vspace{-1em}
        \centering
        \small
        \begin{tabular}{>{\arraybackslash}p{2cm}|>{\centering\arraybackslash}p{1.4cm}>{\centering\arraybackslash}p{1.4cm}>{\centering\arraybackslash}p{1.4cm}}
        \multicolumn{4}{c}{intentions vs tweets type}	\\ \hline
        Intention &	Regular &	Replies &	Quoted  	\\ \hline \hline
        All & 67.4\% & 25.4\% & 7.2\%	\\ \hline \hline
        Converse & 23.5\% & 63.0\% & 13.5\%	\\ \hline
        Promote & 88.3\% & 6.1\% & 5.6\%	\\ \hline
        Share & 81.7\% & 15.1\% & 3.2\% 	\\ \hline
        \textit{NC-UN} & 52.6\% & 39.2\% & 8.2\% 	\\ \hline
        \end{tabular}

        \begin{tabular}{>{\arraybackslash}p{2cm}|>{\centering\arraybackslash}p{0.7cm}>{\centering\arraybackslash}p{0.7cm}>{\centering\arraybackslash}p{0.7cm}>{\centering\arraybackslash}p{0.7cm}>{\centering\arraybackslash}p{0.7cm}}
        \multicolumn{6}{c}{intentions vs. tweets length}	\\ \hline
        Intention &	\textless35 & -70 &	-105 &	-175 &	\textgreater175 	\\ \hline \hline
        All & 13.2\%&23.8\%&22.0\%&22.8\%&18.2\% \\ \hline \hline
        Converse & 11.1\%&12.3\%&21.6\%&25.3\%&29.7\% \\ \hline
        Promote & 9.2\%&27.7\%&23.3\%&21.8\%&18.0\%	\\ \hline
        Share & 11.9\%&28.6\%&24.6\%&25.4\%&9.5\%	\\ \hline
        \textit{NC-UN} & 22.0\%&23.1\%&17.9\%&19.8\%&17.2\%	\\ \hline
        \end{tabular}

        \begin{tabular}{>{\arraybackslash}p{2cm}|>{\centering\arraybackslash}p{1cm}>{\centering\arraybackslash}p{1cm}>{\centering\arraybackslash}p{1cm}>{\centering\arraybackslash}p{1cm}}
        \multicolumn{5}{c}{intentions vs. reactions count}	\\ \hline
        Intention &	0 &	1-5 &	6-10 &	\textgreater 10	\\ \hline  \hline
        All & 67.3\%&24.5\%&5.9\%&2.3\%	\\ \hline \hline
        Converse & 57.4\%&34.6\%&6.8\%&1.2\%	\\ \hline
        Promote &76.7\%&18.9\%&3.2\%&1.2\%	\\ \hline
        Share & 60.3\%&24.6\%&9.5\%&5.6\%	\\ \hline
        \textit{NC-UN} & 64.6\%&26.1\%&6.3\%&3.0\%	\\ \hline
        \end{tabular}
    \label{tab:results-properties}
\end{table}

\subsection{Study 1: Intention Distribution Analysis }
The goal of this study is to investigate the distribution of the intentions for including URLs on sampled tweets using crowd labeling. 
We published a URL-sharing intention annotation task on AMT where we asked participants to select an appropriate intention for including a URL in a tweet.

\textbf{Task Design, Participants, and Study Context.} Inspired by prior work~\cite{snow2008cheap, HettiachchiKG22}, we design clear task instructions and an intuitive annotation interface. We adopt an iterative task refinement process, progressively enriching the information provided to annotators—from tweet text with URLs, to tweet text with the title and textual content of the linked URLs, and finally to the complete tweet context (including hashtags and mentions) together with the title and textual content of the linked URLs. This iterative refinement substantially reduces the annotation error rate from 32.0\% to 16.5\%. In addition, we conduct multiple pilot studies to verify that participants fully understand the task requirements and can reliably assess the intentions underlying URL sharing in tweets.
We require each tweet to be annotated by the same five \textit{eligible} workers to simplify the inter-rater reliability calculations. To ensure that, we published the tweets in a sequence of four batches. Each batch was open for workers who completed the previous batch. The first batch had 100 tweets, and the workers were paid \$0.06 per tweet. We increased the rewards in each following batch by \$0.02 and the number of tweets by 100. As a result of this reward scheme, the entire set was annotated by the same workers. It also sped up the annotation process since the workers were motivated to earn a higher reward. The total cost of the annotation was about \$750 meaning that each worker received \$150 to annotate 1000 tweets.

\begin{figure}[!t]
\centering  
  \includegraphics[width=0.75\columnwidth]{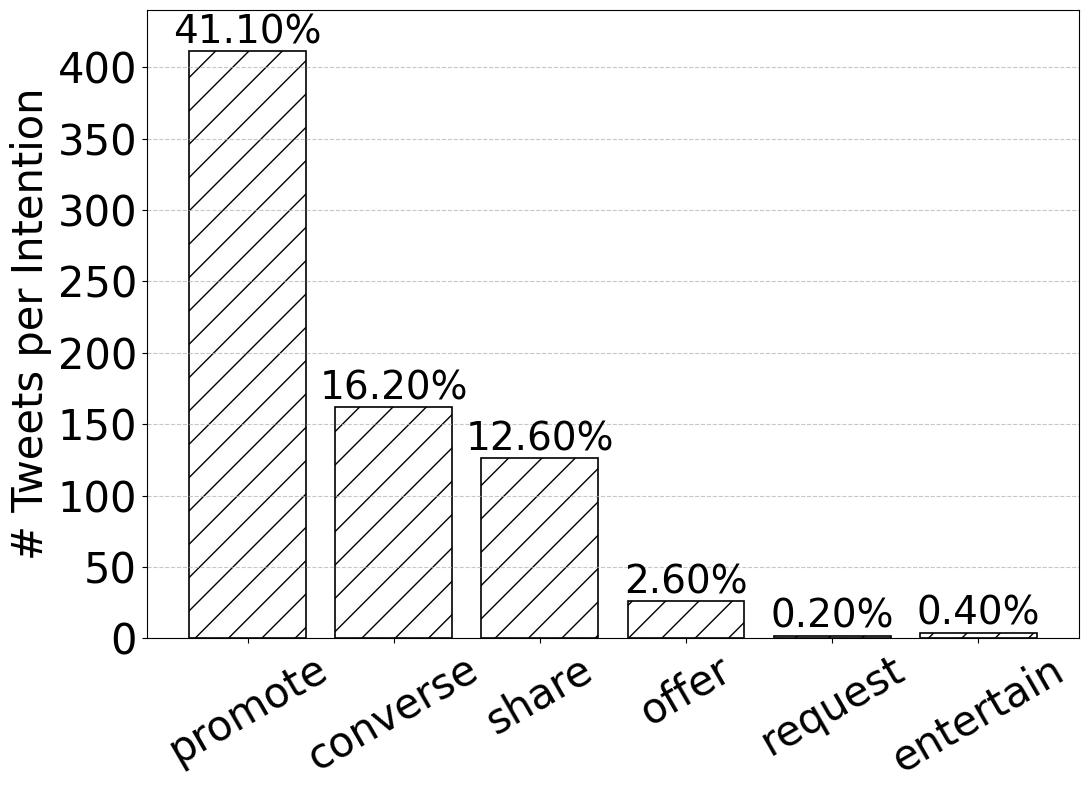}
  \caption{The distribution of overall intentions in Study 1.}
  \label{fig:intentions-pie}
\end{figure}

\textbf{Results.}
On average, workers took 161.8 seconds to annotate a tweet with a standard deviation of 90.1 seconds. The average times of our five workers (in seconds) were 114.1, 125.4, 162.2, 175.3, and 226.9. Naturally, tweets vary in length and complexity. Hence, some workers needed more time to understand and annotate them, which explains the high standard deviation. The resulting annotation is shown in Figure \ref{fig:intentions-pie}. We label the intention of a URL in a tweet based on the \textit{high consensus}. 22.5\% of the tweets with URL have no \textit{high consensus} (we discuss them in Sections \ref{sec:follow_up1}). The most common intentions for tweets with  URLs are \textit{promote}, \textit{converse}, \textit{share}, and \textit{offer}. Less than 1\% of the tweets are annotated with \textit{request} or \textit{entertain} intentions. Finally, the workers give an \textit{uncertain} label for 4.4\% of the tweets with URL. It is worth comparing that to 32.5\%, which is the percentage of the tweets that receive the \textit{uncertain} class using existing taxonomies from related work (Section \ref{sec:gap}). 
This shows that our taxonomy
not only encompasses the known intentions identified in previous studies but also introduces novel categories that contribute to a more nuanced understanding of user intentions in the context of URL sharing.

\textbf{Tweets Properties in Intentions.}
We take a closer look into the distribution of some tweets' properties. We compared these statistics with the ones we reported in Table \ref{tab:results-properties} for the entire dataset and found a few interesting takeaways. For example, 25.4\% of all tweets are replies, but in \textit{converse} and \textit{NC-UN} replies are 63.0\% and 39.2\%, respectively. It is expected in many cases when the URL is intended to be part of the conversation or an argument (i.e., with class \textit{converse}) that the tweet is a reply to other tweets. We also notice that \textit{NC-UN} tweet appeared to have more short tweets ($<$ 35 characters) than the overall tweets (22.0\% vs 12.3\%). In Table \ref{tab:compare-nm}, we focus on comparing tweets that were annotated successfully (with a majority) and tweets with \textit{no intention} or with \textit{uncertain} class (\textit{NC-UN}).

\subsection{Study 2: Effects of Expertise and Context on Annotation Behavior}
The \textit{high consensus} tweets account for 77.5\% of our results, including when the workers agree that the tweet is unclear (\textit{uncertain}). The Fleiss' kappa for all annotation items was $0.216$, and $0.259$ when only considering the \textit{high consensus} tweets. The scores indicate fair agreement between annotators \cite{Landis1977The}. In this study, we explore whether the expertise experience of annotators and more contextual information impact the annotation agreement. 



\textbf{Non-Expert vs. Expert Annotations.}
In addition to the inherent complexity of the task, crowd participants did not receive specialized training nor opportunities to calibrate their judgments with other annotators. Prior studies have shown that, in the absence of domain training, comparing crowdsourced annotations with expert judgments provides an effective means of assessing annotation reliability and task clarity~\cite{snow2008cheap, alonso2013crowdsourcing}. Accordingly, we evaluate the level of agreement between AMT workers and expert annotators. The expert annotations were produced by the authors, who have prior research experience on URL-sharing intentions in social media posts and performed self-guided calibration to ensure consistent intent labeling. We manually annotated a subset of 775 tweets with \textit{high consensus} label from the crowd annotations. The resulting Cohen’s Kappa coefficient is $0.793$, indicating substantial agreement between non-expert and expert annotations.

\textbf{Non-contextual vs. contextual information provided.}\label{sec:follow_up1}
It is reasonable to assume that posts whose intentions are unclear to annotators (\textit{NC-UN}) generally require more contextual information to be accurately judged.
Table \ref{tab:compare-nm} showed that short tweets, replies, or both have more presence of  \textit{NC-UN} tweets (41.3\%, 44.7\%, 48.7\%) than their presence in the overall results (26.9\%). In reply tweets, the workers have no knowledge of the previous conversation. We aim to show that adding context (i.e., a sample of the tweet thread) to the annotation task can help annotators comprehend the tweets better and improve their annotation agreement. Taking advantage of the available context when processing short text has been used in other data annotation problems, such as Named Entity Linking (NEL) \cite{Basile2017Entity}. \citeauthor{Fang2014Entity} employs contextual clues such as the location of a tweet to resolve keyword ambiguity in entity linking in tweets \cite{Fang2014Entity}.

\begin{figure}[!t]
  \centering
    \includegraphics[width=0.9\columnwidth]{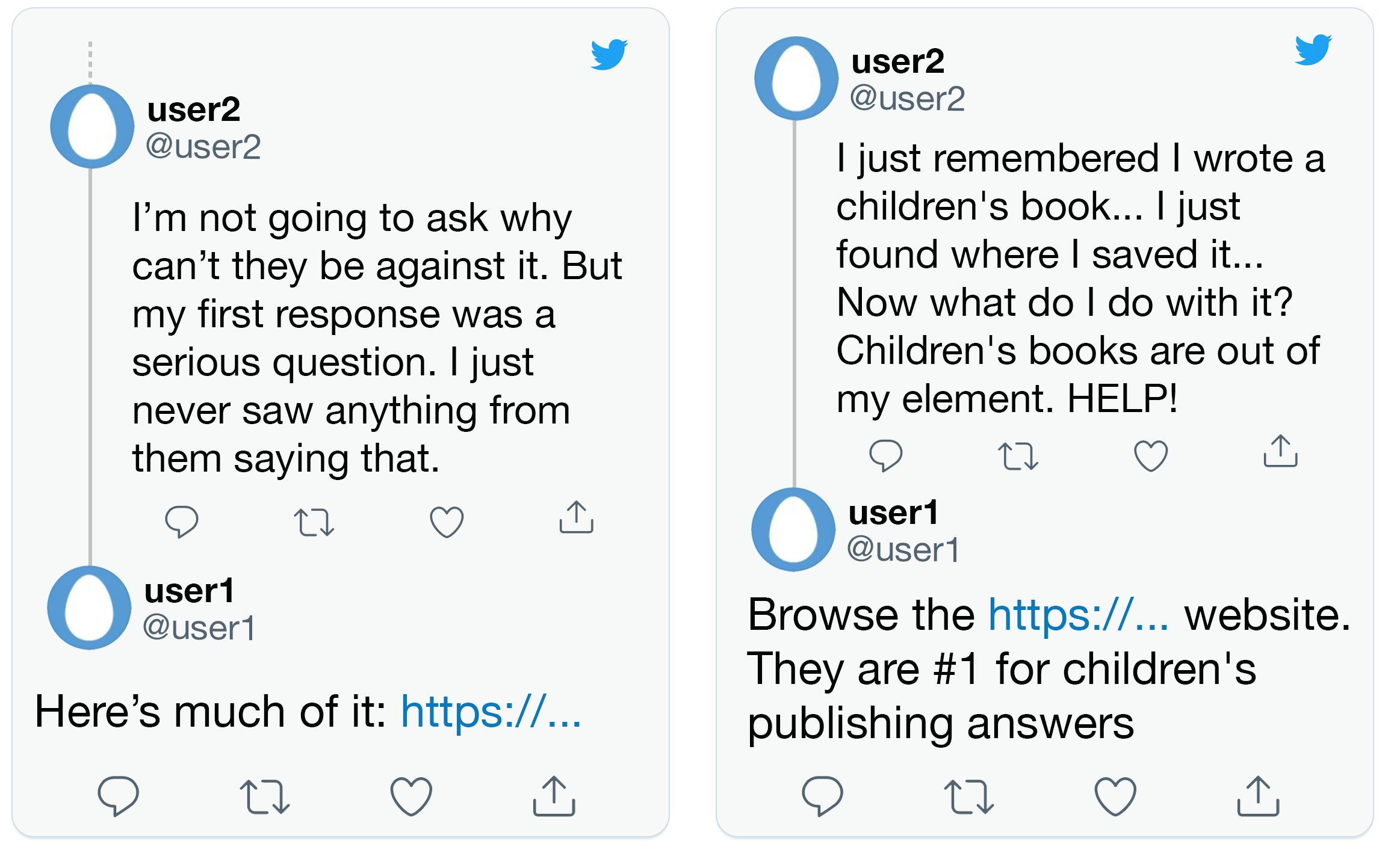}
    \vspace{-1em}
\caption{Example of two tweets where providing context to the workers (conversation) improved their agreement.}
\label{fig:rep1}
\vspace{-10pt}
\end{figure}

In this study, we attempt to improve annotations in reply tweets that are \textit{NC-UN}. We repeat the annotation study with two sets of annotators: (a) the same 5 workers from the main user study and (b) 5 new workers. We provided the annotators with a sample of the conversation (thread) instead of just the standalone tweet. By default, the sample consists of the parent tweet. A user may click and expand the conversation thread with additional tweets, such as replies. We used 83 tweets in this study. Both (a) and (b) produced positive results, 80\%, and 78\% have a majority intention in this study. Figure \ref{fig:rep1} shows two examples: tweet 1 on the left and tweet 2 on the right. In Study 1, the annotations for tweet 1 were [\textit{promote}, \textit{offer} ($\times2$), \textit{share} ($\times2$)] and [\textit{promote} ($\times2$), \textit{offer}, \textit{share} ($\times2$)] for tweet 2. After providing context to the same workers (a) their annotations become [\textit{converse} ($\times3$), \textit{offer} ($\times2$)] for tweet 1 and [\textit{promote}, \textit{offer} ($\times4$)] for tweet 2. These results confirm our assumption that adding context to the tweets helps produce more accurate annotation. We note that the workers in this task need 179.5 seconds on average to complete the annotations. The time increase (from 161.8 seconds) is expected due to the increased workload (i.e., reading the parent tweets in addition to reading the tweet itself).

\begin{table}[!t]
    \centering
    \caption{Comparison between \textit{high consensus} annotations and \textit{NC-UN} in three tweet types: tweets that are replies, have short text ($<$ 35 characters), and tweets that are both.}
    \small
    \begin{tabular}{l|cc}
    \hline
    Property  &	\textit{NC-UN} & \textit{majority} 	\\ \hline \hline
    reply tweets & 41.3\% &58.7\%	\\ \hline
    short tweets & 44.7\% & 55.3\%	\\ \hline
    short + reply tweets & 48.7\% & 51.3\% 	\\ \hline
    \end{tabular}
    
    \label{tab:compare-nm}
\end{table}

\section{Evaluation}
\label{sec:evaluation}

\subsection{Discussion: Intent Taxonomy Comparison}

While prior studies have investigated user intentions in Twitter posts~\cite{Alhadi2011Exploring, GmezAdorno2014Content, Java2007Why}, the intent categories proposed in these works are fully subsumed by our intent taxonomy. Table~\ref{tab:taxonomy_comparison} summarizes the mapping between our taxonomy and those introduced in~\cite{Alhadi2011Exploring, GmezAdorno2014Content, Java2007Why}. A key challenge in performing this mapping is that some prior intent categories lack clear definitions or representative examples~\cite{Alhadi2011Exploring, GmezAdorno2014Content}, making it difficult to accurately align intents based solely on their names. Despite this limitation, the mapping results show that our taxonomy comprehensively covers all previously proposed intent categories. Moreover, our taxonomy introduces an additional intent category, \textit{Entertainment/Humor (Entertain)}, which is absent from existing taxonomies. These findings indicate that our intent taxonomy provides a more comprehensive characterization of user intentions in tweets, applicable to posts both with and without embedded URLs.

\begin{table*}[t]
\centering
\caption{The comparison of intent taxonomies between our intent taxonomy and those proposed in \cite{Alhadi2011Exploring, GmezAdorno2014Content, Java2007Why}}
\vspace{-1em}
\small
\label{tab:taxonomy_comparison}
\begin{tabular}{l|l|l|l}
\hline
 &
  \citeauthor{Alhadi2011Exploring} \cite{Alhadi2011Exploring}  &
  \citeauthor{GmezAdorno2014Content} \cite{GmezAdorno2014Content} &
  \citeauthor{Java2007Why} \cite{Java2007Why} \\ \hline \hline
Information Sharing (Share) &
  Share resources &
  \begin{tabular}[c]{@{}l@{}}News Report (NR)\\      Share Location/Event (SL)\\      Personal Message (PM)\end{tabular} &
  \begin{tabular}[c]{@{}l@{}}Sharing   information/URLs\\      Reporting news\end{tabular} \\ \hline
Entertainment / Humor   (Entertain) &
   &
   &
   \\ \hline
Assistance / Information   Provision (Offer) &
  \begin{tabular}[c]{@{}l@{}}Broadcast   alert/urgent information\\      Express emotions\end{tabular} &
  Question (QU) &
   \\ \hline
Discussion / Opinion   Expression (Converse) &
  Social interaction with people &
  \begin{tabular}[c]{@{}l@{}}News Opinion (NO)\\      General Opinion (GO)\\      Chat (CH)\end{tabular} &
  \begin{tabular}[c]{@{}l@{}}Daily Chatter\\      Conversations\end{tabular} \\ \hline
Promotion / Advertisement   (Promote) &
  Promotion or   marketing &
  Publicity (PU) &
   \\ \hline
Request / Call for Action &
  \begin{tabular}[c]{@{}l@{}}Give or require   feedback\\      Require/raise funding\\      Recruit worker\end{tabular} &
  Question (QU) &
   \\ \hline
\end{tabular}
\end{table*}


\subsection{Example: Intent-aware Microblog Retrieval}
\label{sec:example_retrieval}
In this section, we demonstrate how our intent taxonomy can be leveraged to enhance microblog retrieval through intent-aware reranking and filtering. We examine whether incorporating hyperlink-related intent information—together with the query intent—can improve the ranking quality of retrieval results.



\textbf{Dataset, Model, and Evaluation Metrics.}
We conduct our experiments on the TREC Microblog Track 2011 dataset, using the first 15 queries and their associated tweets. This dataset is widely used for studying microblog search tasks and evaluation methodologies. To clearly illustrate the utility of our intent taxonomy, we adopt BM25~\cite{10.1561/1500000019} as the baseline retrieval model. BM25 is a widely used lexical retrieval method and commonly serves as a first-stage retriever for more sophisticated reranking models. Our intent-aware approach augments the baseline ranking by incorporating intent features derived from the hyperlinks embedded in tweets. We evaluate retrieval effectiveness using standard ranking metrics, including nDCG@10 and MAP.

\textbf{Experiment Design.}
\textit{Baseline (BM25)}: given a query and a collection of microblog posts (i.e., tweets), we apply BM25 to retrieve the top-50 candidate tweets.
\textit{Intent-aware reranking (BM25 + Intent)}: for tweets within the top-50 results that contain hyperlinks, we annotate each tweet with the most appropriate intent category using our taxonomy. We also infer the intent of the query using the same taxonomy. We then augment the original query by concatenating it with the inferred query intent, and similarly augment each tweet by concatenating its text with the corresponding URL-sharing intent. Finally, we rerank the top-50 candidate tweets using BM25 over the augmented representations. 


\textbf{Results.}
The experimental results are reported in Table~\ref{tab:rank_performance}. We compare the baseline BM25 model with an intent-aware reranking variant (BM25+Intent), which incorporates the annotated hyperlink intent as an additional feature. The results indicate that explicitly modeling the intent behind hyperlink inclusion leads to consistent improvements in retrieval performance, demonstrating the effectiveness of intent-aware ranking in microblog search.

Beyond reranking, intent annotations also enable filtering of retrieved tweets whose intents are misaligned with the query intent. For example, consider the query \textit{``thorpe return in 2012 olympics''}. Using the query intent taxonomy proposed in~\cite{10.1145/1753846.1754140}, we classify this query as \textit{Informational/Undirected/Closed (I/U/C)}, indicating a user intent to obtain factual information about a specific topic. Most of the top-ranked tweets retrieved by BM25 that contain hyperlinks are annotated as \textit{Information Sharing}, which aligns well with the query intent. However, one retrieved tweet—\textit{``clue : arrested development edition : sure there's always money in the banana stand -- but where's the murder wea [Link]''}—is labeled as \textit{Entertainment/Humor}. Although this tweet shares lexical overlap with the query, inspection of the linked content confirms that it is primarily humorous and does not satisfy the informational intent of the query. This example illustrates how intent-aware filtering can help remove semantically mismatched results that are otherwise ranked highly by lexical similarity alone.




\begin{table}[]
\centering
\caption{TREC ranking performance.}
\vspace{-1em}
\small
\label{tab:rank_performance}
\begin{tabular}{l|l|l}
\hline 
              & nDCG@10 & MAP    \\ \hline \hline
BM25          & 0.4166  & 0.4518 \\ \hline
BM25 + intent & 0.4374    & 0.4757 \\
\hline 
\end{tabular}
\vspace{-1em}
\end{table}

\section{Related Work}\label{sec:related_work}
Communication researchers have developed a long list of why people communicate and much empirical work has been completed to assess the relative effectiveness of face-to-face, various forms of mediated-interpersonal (e.g., e-mail, phone), and a wide range of mass media channels concerning these functions. \citeauthor{Flanagin2001Internet} provide an exhaustive list of 21 reasons why people engage in various types of communication, and they find these 21 individual causes form nine clusters \cite{Flanagin2001Internet}. These clusters are as follows: information, learning, play, leisure, persuasion, social bonding, relationship maintenance, problem-solving, status, and insight. While the uses and gratifications research tradition has focused primarily on individuals as media consumers, it has been used to study why people use and what they get out of various social media experiences (e.g., \cite{Chen2011Tweet, Phua2017Uses, Smock2011Facebook}). Social media platforms allow individuals not only to be message consumers but also content creators. With this shift comes asking not just what information you get from social media sources, but what are the motivations behind and needs for creating social media content.

We enumerate several problems where intent was the main scope of the study. The \textit{query intent} classification aims to learn what the user is searching for by classifying search engine queries \cite{Hu2009Context, Hu2009Understanding, Li2010Understanding, Li2008Learning}. They generally utilize user click information to classify user intent in queries. Another research direction is that of discovering online commercial intention \cite{Dai2006Detecting, Hu2009Context, Hu2009Understanding}. This problem is considered a user intention problem since it relies on user queries and web browsing history to detect the user's commercial intent \cite{Dai2006Detecting}.

In recent years, research has pivoted toward identifying user intention in \textit{computer-mediated communication}, such as discussion forums, mobile phone text messages, and social media posts \cite{Java2007Why, Chen2013Identifying, Alhadi2011Exploring, Purohit2015Intent}. \citeauthor{Java2007Why} studied users' behaviors and intentions on Twitter and concluded that there are four main types of user intentions on Twitter: daily chatter, conversations, sharing information, and reporting news \cite{Java2007Why}. They analyzed these intentions on both user and community levels and categorized relations between users into information seekers/sources and friends \cite{Java2007Why}. \citeauthor{Alhadi2011Exploring} found that a user's intent usually falls into one of eight intention categories, which include promoting, social reaction, and expressing emotions \cite{Alhadi2011Exploring}. \citeauthor{Chen2013Identifying} employed a transfer learning method to perform binary classification for \textit{intent in discussion forums} \cite{Chen2013Identifying}. Other studies on intentions on Twitter were centered on topical tweets, such as tweets about crisis events or elections \cite{Purohit2015Intent, Mohammad2015Sentiment}.  \citeauthor{Holgate2018Why} studied the \textit{intent behind using swear words} in tweets and presented a novel dataset of 7800 tweets that are manually labeled with one of six categories of vulgar word intentions \cite{Holgate2018Why}. 

While most of those works are about the intentions in user's text, others investigated the use of other elements, such as \textit{emoticons and emojis} \cite{Derks2008Emoticons, Tauch2016The, Hu2017Spice}. \citeauthor{Derks2008Emoticons} studied the role of emoticons in chat messages and reported that these three intentions are the most common intentions for using emoticons expressing humor or emotion and strengthening these expressions \cite{Derks2008Emoticons}. The work of \citeauthor{Hu2017Spice} studied the intentions and sentiment effects of using emojis on Twitter and reported the results of two user studies on these aspects \cite{Hu2017Spice}. Their results showed that expressing sentiment and strengthening the expression are the two most common intentions of using emojis on Twitter \cite{Hu2017Spice}. \citeauthor{SalahEldeen2013Reading} study the relationship between temporal inconsistency and the author's intentions in posting links on Facebook and Twitter \cite{SalahEldeen2013Reading}.
In other words, they investigate whether the author intends to convey the current content of the shared web page or not \cite{SalahEldeen2013Reading}.
 In another work by \citeauthor{Holton2014Seeking}, they surveyed Twitter users and asked them about their sharing URL habits and what motivates them to do so \cite{Holton2014Seeking}. This work differs from ours because of the distinction between motivation and intention when it comes to human behavior. In psychology, human intentions explain why someone plans to do something, while motivation can look back at reasons for past actions or categorize actions within a broader framework \cite{kenny, scheer}. Whereas \citeauthor{Holton2014Seeking} inquire about users' overall motivation for the habit of sharing URLs, our investigation focuses on discerning the specific intention behind sharing a URL in an individual tweet.


\section{Future Work}
Our future work includes a couple of directions: enhancing (microblog) information retrieval and integrating insights from communication science. In the former, we plan to extend the observations in Section~\ref{sec:NLPIR} by embedding our taxonomy into a microblog retrieval system \cite{Cambazoglu2021An, li2023intentawarerankingensemblepersonalized}, enabling intention-aware ranking and improved relevance for tasks such as credibility assessment and misinformation detection. 
In the latter direction, we aim to link our computational approach with theoretical perspectives from communication science, where communication intentions are central to understanding message effects. For example, research based on Reactance Theory investigates how audiences respond to perceived persuasive intent \cite{Quick2015Prospect,Dillard2005On}, while studies on perceived partisan bias examine when audiences interpret news as persuasion rather than information \cite{Eveland2003The,Morris2007Slanted}. These traditions highlight how intention distinguishes persuasion, social influence, and manipulation.
Integrating such theoretical insights with computational modeling offers a promising interdisciplinary path for studying and modeling communicative intent in online discourse \cite{Alshehri21,Alshehri23}.

\section{Conclusion}

Understanding the intents of including URLs in posts provides an essential foundation for information retrieval on social media platforms, where URLs act as bridges connecting user-generated posts to the broader web. To systematically capture and model these purposes, we developed an intent taxonomy for including URLs in posts through a hybrid construction method that combines large-scale crowdsourced annotation with LLM assistance and conducted two studies examining intention distribution and the effects of annotator expertise and contextual information on annotation agreement. Finally, we compare our intent taxonomy with those proposed in prior research and demonstrate a potential application. Overall, our work offers a reproducible model for building empirically grounded, LLM-augmented taxonomies applicable to future IR and NLP research.

\begin{acks}
This work was supported in part by the U.S. National Science Foundation awards III-2107213 and 2026513.
\end{acks}

\bibliographystyle{ACM-Reference-Format}
\bibliography{all}

\end{document}